\documentstyle[preprint,revtex]{aps}
\tightenlines

\def\lag{\langle}
\def\rag{\rangle}

\def\omegac{\omega_{\rm c}}
\def\Ne{N_{\rm e}}

\def\me{m_{\rm e}}
\def\omegaz{\omega_0}

\def\Vc{V_{\rm c}}
\begin{document}
%\draft
\begin{title}
Fractional Quantum Hall States in Narrow Channels\\
\end{title}
\author{Daijiro Yoshioka}
\begin{instit}
Institute of Physics, University of Tokyo\\
Komaba, Meguro-ku, Tokyo 153 Japan\\
\end{instit}
\receipt{7 December 1992}

\begin{abstract}
A model system is considered where two dimensional
electrons are confined by a harmonic potential in one direction,
and are free in the other direction.
Ground state in strong magnetic fields is investigated through
numerical diagonalization of the Hamiltonian.
It is shown that the fractional quantum Hall states are realized
even in the presence of the external potential under suitable
conditions, and a phase diagram is obtained.
\end{abstract}
\pacs{}

\narrowtext

%\section{}
%\label{sec:level1}

It has been known that two-dimensional electron system in a
strong magnetic field condenses into an incompressible liquid at low
temperature which is known as Laughlin state.\cite{laugh}
This state causes the fractional quantum Hall effect (FQHE).\cite{fqhe}
In the related phenomenon, integer quantum Hall effect (IQHE),
importance of the edge states has been recognized.\cite{edge}
Thus it is also of interest to know the details of the edge
in the FQH states.
Since
effect of the edge should be dominant
in the narrow channel,
it is desirable to know the properties of the ground state in
quantum wire in a situation where FQHE is observed for a bulk
systems.

Chui\cite{chui} has considered narrow
channel system in a strong magnetic field.
He has taken a hard wall boundary condition.
He found a transition between the FQH state and $2k_F$ CDW state
as the channel width is varied.
In the present letter we consider a system with different boundary
conditions.
In our model the electrons are confined by a parabolic potential
in the lateral direction.
Thus the width of the system is determined spontaneously:
It depends on the electron density per unit length of the channel,
and on the strength of the confining potential relative to the interaction
between electrons.
We expect that as the confining potential gets weaker the electrons
expand laterally, and in the course of the expansion different ground states
are realized successively which include the FQH states with various
filling factors.
In a preliminary work\cite{dy} we used model interaction potential\cite{hal}
to investigate the ground state.
The model potential is such that the Laughlin state at 1/3-filling
becomes the ground state in the limit of strong interaction.
There we investigated how the ground state evolves from a filled
Fermi sea in the limit of vanishing interaction
to the 1/3 state in the opposite limit.
In this letter we use the realistic Coulomb interaction.

A model related to the present one, where interacting electrons are
confined by two-dimensional harmonic potential, has been
investigated.\cite{kin}
This is a model for the quantum dot.
In this case it was found that 2/3 Laughlin state is realized
at the central part of the system  under suitable conditions.
Our model has the similar aspect to this model.
We investigate if such a state is realized also in our quantum wire
or not.

Another remarkable aspect of our model is that
since it is  quasi-one-dimensional,
it is possible to consider it as a strongly correlated one dimensional
system.
Single particle part of our Hamiltonian has the parabolic dispersion,
which is nothing but that of the one-dimensional electrons
in a continuous space.
The only difference from the genuine 1-d system is the interaction
part of the Hamiltonian:
Our system has slightly complicated form, although the difference
may not be so important.
Thus this is a good model to study one dimensional highly
correlated electron systems.

We consider two-dimensional electrons on the $xy$-plane
in an external confining potential.
The potential is flat in the $x$-direction,
but parabolic in the $y$-direction.
The length of the system in the $x$-direction is $L_x$, and
we impose periodic or antiperiodic boundary condition in this direction.
A strong magnetic field, $B$, is applied in the $z$-direction.
Thus the single particle part of the Hamiltonian is given by
\begin{equation}
H_0 =  {{1}\over{2\me}} [(p_{x} + eBy)^2 + p_{y}^2]
+ {1\over 2}\me\omegaz^2y^2.\label{one}
\end{equation}
Here $\me$ is the mass of the electrons, $\omegaz$ gives the strength of
the confining potential.
The eigenstate of this Hamiltonian is easily obtained to be
\begin{equation}
\psi_{k_x,n}({\bf r}) = \exp({\rm i}k_xx-{{{\tilde y}^2}\over{2\lambda^2}})
{\rm H}_n({{\tilde y}\over{\lambda}}),
\label{two}
\end{equation}
where $\lambda = \sqrt{\hbar/\me\Omega}$
is the effective Larmor radius,  $\Omega = \sqrt{\omegaz^2 +
\omegac^2}$ and $\omegac = eB/\me$.
The wave function is localized in the $y$-direction around $\tilde y=0$,
where $\tilde y = y+ {{(\hbar k_x\omegac)}/{(\me\Omega^2)}}$.
The wave number in the $x$-direction is quantized to be
$k_x = (2\pi/L_x)m$,
with quantum number $m$ being an integer (half odd integer) under
the periodic (antiperiodic) boundary condition.
The momentum $\hbar k_x$, or $m$, determines the center coordinate
of the wave function in the $y$ direction.
The function ${\rm H}_n$ is the Hermite polynomial.

This state has the energy,
$
E_{k_x,n} = (n + {1/2})\hbar\Omega +{({\hbar^2k_x^2\omegaz^2})
/({2\me}{\Omega^2})}.
$
Hereafter we assume that the magnetic field is strong enough
that we are allowed to consider
only the states with $n=0$.
Thus our system has a dispersion the same as free 1-d electrons with
effective mass which can be quite heavy: $(\Omega^2/\omegaz^2)\me$.
We introduce creation (annihilation) operator $a_m^\dagger (a_m)$
corresponding to the wave function
$\psi_{k_x,n}({\bf r})$ with $n=0$ and $k_x=2\pi m/L_x$.
Then the second quantized single electron Hamiltonian is written as
follows except for a constant term:
\begin{equation}
H_0 =  \sum_m {{\hbar^2k_x^2}\over{2\me}}{{\omegaz^2}\over{\Omega^2}}
a_m^\dagger a_m.
\label{four}
\end{equation}

Interaction between electrons is the usual Coulomb interaction:
$V({\bf r}) =e^2/4\pi\epsilon r$.
Considering the boundary condition in the $x$-direction,
we truncate the potential such that $V(x,y)=0$ for $|x|>L_x/2$.
Now the interaction part of the Hamiltonian is given as
\begin{equation}
H_{\rm int} = {1\over 2}\sum_m \sum_{m_p} \sum_{m_q} f(m_p,m_q)
a_{m+m_q}^\dagger a_{m-m_p}^\dagger a_{m-m_p+m_q} a_m,
\label{five}
\end{equation}
where
\begin{eqnarray}
f(m_p,m_q) = {{1}\over{\sqrt{2\pi}\lambda L_x}}
&&\int_{-L_x/2}^{L_x/2} {\rm d}x \int_{-\infty}^{\infty} {\rm d}y
V(x,y)\nonumber\\
&&\times \exp [ - {{2\pi{\rm i}}\over{L_x}} m_q x -
{{1}\over{2\lambda^2}}(y+\alpha m_p\lambda)^2 -
{{\alpha^2m_q^2}\over{2}} ],
\label{six}
\end{eqnarray}
with $\alpha \equiv L_x/2\pi\lambda$.
It should be noticed that for ordinary 1-d electrons the matrix element
$f(m_p,m_q)$ does not  depend on $m_p$.

Thus in the present model there are three dimensionless parameters
except for number of electrons in the system $\Ne$,
namely $\alpha$, $\omegac/\Omega$,
and $\Vc /E_0$.
Here
$\Vc = (e^2/4\pi\epsilon\ell)$ gives the measure of the Coulomb interaction
and $E_0 = (\hbar^2/2\me\lambda^2)(\omegaz^2/\Omega^2)$
has been chosen as the unit of energy.
Among them we will set $\omegac/\Omega=1$ in the following calculation
assuming $\omegac >> \omegaz$.
\par

When we change
$\Vc /E_0$, various  ground states
are realized.
Here we focus our attention to where FQH-states are realized.
The analytic form of Laughlin's FQH state in this gauge
has been written down by
Thouless.\cite{thouless}
However, here we construct the $1/p$ FQH state numerically as the ground state
in the model potential.\cite{dy}
Using electron-hole symmetry we can also construct $\nu=2/3$ state
for systems with even number of electrons.\cite{footnote}
We calculate the overlap between the true ground state and these
model states to investigate where in the parameter space these
FQH states are realized.

In order to obtain the ground state we impose the periodic (antiperiodic)
boundary condition in the $x$-direction, when the number of the
electrons is even (odd).
Then the ground state is realized among the states with zero total momentum
in the $x$-direction.
To perform the numerical calculation we need to restrict the total number
of the single electron states.
We take into account sufficient number of states that the amplitude of the
occupation of the highest energy single electron state is less than
$10^{-4}$.
In spite of this restriction, it is difficult to investigate a system
near $\nu=1/5$, if the electron number exceeds seven.
Therefore we consider system with 6 electrons in this letter.
In this case we can discuss overlap of the ground state with
the $2/3$, $1/3$ and $1/5$ FQH states.

We investigated the ground state at $1.0 < \alpha < 2.5$ and
$0 < \Vc/E_0 < 200.$
We have chosen only moderate value for $\alpha$, since
too large or too small $\alpha$ makes the aspect ratio of the system
too large:
If $\alpha$ is too large, the length of the system $L_x$
is much larger than the width.
The system is truly one-dimensional, and the electrons line up
to form $2k_{\rm F}$-CDW state, or the Wigner solid.
On the other hand, if $\alpha$ is too small, the width of
the system becomes much larger than the length.
In this case electrons line up again, but in the lateral direction
to form the Wigner solid.
Then our system ceases to be a good model for real narrow channel system.
The aspect ratio becomes unity for the three model states of $\nu= 2/3$,
1/3 and 1/5 at $\alpha=1.26$, 1.60 and 2.03, respectively.

To illustrate how the ground state evolves we show in Fig.1(a)
the overlap of the ground state with the model states,
and ground state expectation values of the components of the Hamiltonian,
$\lag H_0 \rag/E_0$ and $\lag H_{\rm int} \rag/E_0$ at $\alpha = 1.5$
(Fig.1(b)).
The ground state changes almost abruptly from one state to the other,
when $\Vc /E_0$ is changed.
Exactly speaking this is not a first order transition.
Because these states have the same symmetry, the character of the
ground state changes continuously from one ground state to the other
at the boundary.
However, this is a finite size effect.
We expect that
the transition becomes sharp for an infinitely long system.

We repeat the calculation changing the value of $\alpha$ to obtain
a phase diagram shown in Fig.2.
In this range of $\alpha$ there is a well defined phase boundary
for the 1/3 FQH state.
However, for 1/5 FQH state the boundary becomes obscured, when $\alpha$
becomes small ($<1.7$), since there the width of the system becomes much larger
than the length $L_x$.
Boundary for the 2/3 state also becomes less clear when $\alpha$ becomes
larger than 2, where the system becomes one-dimensional.

We have seen that for suitable choice of parameters Laughlin type
FQHE states are  good approximation to the ground state even
in the present system, where the effect of edge should be
important.
Thus the present system is a good model to study the details of
edges in the Laughlin state, which we will pursue in the future
investigation.
Here we remark that reconstruction of edges as discussed in
\cite{fer} does not occur as evident from the large overlap
with the FQH states.
The most important origin of this difference is the difference in the
boundary condition:
Their confinement potential is flat at the bottom and linear at the edges.
It is conceivable that their use of the Hartree Fock approximation
also causes the difference.

The Laughlin state in the present geometry
has a sharp boundary for the occupation
of the single electron state:
only the lowest $(\Ne-1)p+1$ states are occupied in the $1/p$ FQH
state with $\Ne$ electrons.
Namely in the language of the one-dimensional system there is a
discontinuity in the momentum distribution at the Fermi points.
Thus this state is distinct from the Tomonaga-Luttinger liquid\cite{tl}
in two ways: (i) the Luttinger sum rule\cite{lutt} is not satisfied
and (ii) discontinuity at the Fermi points.
On the other hand the true ground state with the Coulomb interaction
is slightly different from the Laughlin state.
Occupation of the states outside of the Fermi points are quite small
but not zero.
Unfortunately our system is too small to examine if the momentum
distribution has a discontinuity or not.
However, it is quite likely that for an infinitely long system
the momentum distribution has no discontinuity like
the Tomonaga-Luttinger liquid.

In the present letter we have not discussed the ground states
realized between the Laughlin states.
This is because of the difficulty to characterize it.
In some region of the parameter space it looks like a CDW state
as Chui has discussed.
We leave the characterization to a future investigation.

We expect the present system  with
the values of parameters considered here can be realized
actually.
If experiments will be done in a magnetic field of the order of
10T,
ordinary two-dimensional electron system confined laterally
into width of about 0.1$\mu$m will fall into the parameter space
in Fig.1.
The strength of the potential $\omegaz$ depends on how the confinement
is realized.
If we adopt a value of $\omegaz=100K$ for example, typical of the
quantum dot system,
$V_c/E_0 \simeq 30$ for GaAs-AlGaAs, which is suitable for the
1/3 state to be realized.
Thus remaining problem is how to detect the realization of
the Laughlin state.
Unfortunately  we cannot expect quantization of the Hall resistance
in such a narrow system.
However, transport of electrons along the channel should depend on the
ground state, and it will be possible to distinguish the ground states.

This work is supported by Grant-in-Aid for Scientific Research on
Priority Areas ^^ ^^ Computational Physics as a New Frontier in
Condensed Matter Research" (04231105) and by
% Grant-in-Aid for Scientific Research on Priority Areas
% ``Science of High Tc Superconductivity" (04240214)
Grant-in-Aid for Scientific Research (04640361)
from the Ministry of Education,
Science and Culture.

\figure{
(a) The overlap of the ground state with the 2/3 and 1/3 FQH states, and
(b) the expectation values of $H_0$, eq.(\ref{four}),
and $H_{\rm int}$, eq.(\ref{five}), in the
ground state
are shown as a function of $\Vc /E_0$
for a system with  $\alpha=1.5$.\label{fig1}}

\figure{Phase diagram of the present system.
The regions where the $\nu=1$, 2/3, 1/3 and 1/5 FQH states are
realized are shown
on the parameter space spanned by $\alpha=L_x/2\pi\ell$
and $\Vc /E_0$.\label{fig2}}

\begin{references}
\bibitem{footnote} In the $1/3$ FQH state $\Ne$ electrons occupy the
lowest $3\Ne-2$ states.
We consider this state as the $2/3$ FQH state where $2(\Ne-1)$
holes occupy the lowest $3\Ne-2$ states.
\end{references}
\end{document}